\documentclass[%
 reprint,
 amsmath,amssymb,
 aps,
]{revtex4-2}

\usepackage{graphicx}
\usepackage{dcolumn}
\usepackage{bm}
\usepackage{braket} 
\usepackage{xcolor}
\begin{document}

\preprint{APS/123-QED}

\title{Unique Gravitational-Wave Signals from Negative-Mass Binaries}

\author{Oem Trivedi$^{1}$, Abraham Loeb$^{2}$}
\affiliation{$^{1}$Department of Physics and Astronomy, Vanderbilt University, Nashville, TN, 37235, USA}
\affiliation{$^{2}$Astronomy Department, Harvard University, 60 Garden St., Cambridge, MA 02138, USA}

\email{Email: oem.trivedi@vanderbilt.edu \\ Email: aloeb@cfa.harvard.edu}

\date{\today}

\begin{abstract}
Negative masses have long been explored, but their observational viability remains unclear. In this work, we develop a unified, observationally testable framework to constrain negative masses using both coupling level and dynamical probes. We establish that while dipole radiation bounds require universality of gravitational charge, the intrinsic dynamics of negative mass binaries generically lead to anomalous behaviors such as anti-chirps, dispersal and runaway motion. These signatures are absent in current gravitational wave observations, providing a robust exclusion channel independent of modified gravity assumptions.

\end{abstract}

\maketitle

Negative masses have been a topic of interest in gravitational physics for quite some time. Starting with the foundational work of Bondi in 1957 \cite{bondi1957negative}, it was realized that systems involving positive and negative masses lead to highly non-trivial dynamics, including runaway motion and unconventional gravitational interactions. Since then, negative masses have been explored across a wide range of contexts, including their role in general relativity, exotic compact objects and black hole physics, as well as in cosmological settings where they have been invoked in attempts to explain dark energy like behavior or to construct alternative cosmological models such as the Bondi universe \cite{nm1manfredi2026bondi,nm2nojiri2026may,nm3farnes2018unifying,nm4belletete2013negative,nm5bonnor1989negative,nm13cebeci2006negative,nm14mann1997black,forward1990negative}. More broadly, they have appeared in modified gravity frameworks and discussions of exotic matter, where they challenge conventional assumptions about energy conditions and gravitational coupling \cite{nm6moller2017quantum,nm7khamehchi2017negative,nm9socas2019can,nm10petit2014negative,nm11najera2021negative,nm12manfredi2018cosmological}. \\

The peculiar dynamical behavior that negative masses exhibit such as runaway acceleration and apparent violations of intuitive energy considerations, raise serious concerns about their physical viability. Moreover, despite decades of theoretical exploration, there has been no direct observational evidence supporting their existence and this motivates the need for a systematic framework that goes beyond purely theoretical arguments. Instead one requires robust, observationally testable constraints to really probe negative masses properly. The aim of this Letter is to put viable, observationally testable constraints on negative masses and provide a strong criterion which can either rule out this paradigm completely or guide us better under which regimes one could observe such negative mass objects. \\

Dipole radiation provides some of the cleanest constraints on deviations from universal gravitational coupling \cite{dip1boundwang2022simultaneous,dip2lazaridis2009generic,dip3will1977gravitational,dip4krisher1985dipole,dip5barausse2016theory,dip6vasuth2003gravitational,dip7weisberg1981gravitational,dip8arun2012generic}. It is so because in a general framework one may define a channel-dependent gravitational charge $q^X$ and corresponding charge to mass ratio is given by
\begin{equation}
\alpha^X = \frac{q^X}{m}.
\end{equation}
Before proceeding, it is useful to clarify the mass conventions used in this work. In general, one may distinguish inertial mass, which controls the response of an object to an applied force, active gravitational mass, which sources the gravitational field and passive gravitational mass, which controls the response to an external gravitational field. The equality of active and passive gravitational mass is required for momentum conservation. Since momentum conservation is associated with spatial translation symmetry, it is not optional for a consistent physical theory. We therefore identify active and passive gravitational mass in what follows and refer to this common quantity as the gravitational mass or, more generally the gravitational charge. By contrast, the equality between inertial mass and gravitational mass is the content of the equivalence principle, which has been experimentally verified for ordinary positive mass matter but not for any hypothetical negative mass sector. We hence use $q^X$ as a generalized, channel dependent gravitational charge in an observational channel $X$, while $m$ denotes the inertial mass entering the charge-to-mass ratio $\alpha^X=q^X/m$. The dipole-radiation analysis later constrains possible deviations in this ratio, whereas the later quadrupole waveform analysis assumes the stronger signed mass realization in which inertial and gravitational masses remain equal even for negative masses. For two bodies in a binary system labeled $1$ and $2$, the difference in their effective couplings in a given channel $X$ is
\begin{equation}
\Delta \alpha^X = \alpha_1^X - \alpha_2^X.
\end{equation}
In theories that admit additional radiative degrees of freedom beyond the tensor modes of general relativity such as scalar tensor theories, the leading correction to gravitational radiation appears as a dipole contribution. We can then write the energy loss due to gravitational radiation as
\begin{equation}
\dot{E} = \dot{E}_{\rm GR} \left[1 + B \left(\frac{M}{r}\right)^{-1} \right],
\end{equation}
where $\dot{E}_{\rm GR}$ is the standard quadrupole formula contribution, $M$ is the total mass of the system, and $B$ parametrizes the strength of dipole radiation. $B$ is directly related to the difference in gravitational charge to mass ratios of the two bodies in such frameworks, with this given as
\begin{equation}
B = \frac{5}{96} (\Delta \alpha^{\rm dip})^2.
\end{equation}
Observational constraints from binary pulsars and gravitational wave inspirals require $B$ to be extremely small, typically $B \lesssim 10^{-7}$ \cite{dip1boundwang2022simultaneous} and this would mean that
\begin{equation}
|\Delta \alpha^{\rm dip}| \lesssim \sqrt{\frac{96 B}{5}},
\end{equation}
this means that the effective gravitational coupling in the dipole channel must be nearly identical for all objects in a binary system. It is important to emphasize that this constraint does not exclude the existence of negative mass objects per se, but does something deeper. It constrains the difference in gravitational charge to mass ratios. in particular, a scenario in which positive and negative masses carry opposite gravitational charges such that
\begin{equation}
\left(\frac{q}{m}\right)_+ = +1, \quad \left(\frac{q}{m}\right)_- = -1,
\end{equation}
would give
\begin{equation}
\Delta \alpha^{\rm dip} = 2,
\end{equation}
and therefore
\begin{equation}
B = \frac{5}{96} (2)^2 \sim 0.21,
\end{equation}
which is many orders of magnitude larger than observational bounds. Such a configuration is therefore ruled out but if negative mass objects satisfy
\begin{equation}
\left(\frac{q}{m}\right)_- \simeq \left(\frac{q}{m}\right)_+,
\end{equation}
then $\Delta \alpha^{\rm dip} \simeq 0$ and dipole radiation constraints are satisfied. Thus, dipole radiation excludes negative masses with non-universal gravitational charge, but does not exclude negative masses with universal coupling. Note that we get dipole gravitational radiation after introducing negative masses for the same reason we get dipole electromagnetic radiation, as we now have both positive and negative gravitational “charges”.
\\
\\
The dipole channel, however, represents only one aspect of gravitational interaction and one can think of more general scenarios as well. In a more general setting, different physical probes are sensitive to different effective couplings and we may therefore introduce a set of channel dependent charge to mass ratios which are
\begin{equation}
\alpha^X = \frac{q^X}{m},
\end{equation}
where $X$ labels different physical processes such as conservative orbital dynamics, radiative energy loss, gravitational lensing, propagation of gravitational waves and cosmological growth \cite{fut1yagi2016challenging,fut2barausse2015gravitation,fut3barausse2017testing,fut4das2025testing,fut5lambiase2021constraints}. For each such channel, one can define
\begin{equation}
\Delta \alpha^X = \left(\frac{q^X}{m}\right)_+ - \left(\frac{q^X}{m}\right)_-.
\end{equation}
Each observational probe then constrains the corresponding $\Delta \alpha^X$ to lie within some small tolerance $\epsilon_X$
\begin{equation}
|\Delta \alpha^X| \le \epsilon_X.
\end{equation}
For example, in the conservative sector we see that the force law between two bodies can be written as
\begin{equation}
F = -\frac{G m_1 m_2}{r^2} \alpha_1^{\rm cons} \alpha_2^{\rm cons},
\end{equation}
so that any difference in $\alpha^{\rm cons}$ between objects would lead to violations of the equivalence principle and anomalous orbital dynamics. Also in gravitational lensing the deflection angle can be expressed as
\begin{equation}
\hat{\alpha}_{\rm lens} = \alpha^{\rm lens} \frac{4 G M}{c^2 b},
\end{equation}
and deviations in $\alpha^{\rm lens}$ would alter observed lensing signatures while in cosmology, the growth of structure depends on an effective gravitational coupling
\begin{equation}
\ddot{\delta} + 2 H \dot{\delta} = 4 \pi G \alpha^{\rm cosmo} \rho \delta,
\end{equation}
so that differences in $\alpha^{\rm cosmo}$ would modify large scale structure formation.
\\
\\
This motivates a generalized framework in which gravitational charge is not a single quantity, but is instead a set of effective couplings across different channels. Requirement of consistency with observations is then that these couplings must be universal across all types of matter, including any hypothetical negative mass sector and this requires
\begin{equation}
\Delta \alpha^X = 0 \quad \text{for all channels } X,
\end{equation}
within observational uncertainties. The exclusion criterion for negative mass models can therefore be stated as follows: If there exists any channel $X$ for which
\begin{equation}
|\Delta \alpha^X| > \epsilon_X,
\end{equation}
then the model is ruled out by the corresponding probe and conversely a negative mass sector can evade current constraints only if it satisfies
\begin{equation}
\left(\frac{q^X}{m}\right)_- \simeq \left(\frac{q^X}{m}\right)_+
\end{equation}
for all observable channels $X$. This establishes that the viability of negative masses is not determined by the sign of the mass alone, but by the universality of their gravitational response across all experimentally accessible probes.
\\

Gravitational waveform constraints provide us an independent and particularly powerful avenue for probing the existence of negative masses. Unlike dipole radiation constraints which depend on the presence of additional radiative channels and non-universal gravitational couplings, waveform constraints arise already within the standard quadrupole emission framework of general relativity \cite{lvk1abbott2019tests,lvk2abbott2021tests,lvk3ghosh2018testing,lvk4yunes2016theoretical}. They do not,hence, rely on modifications to gravity but instead follow directly from the dynamics of systems containing negative masses.
\\
\\
We start by considering a binary system with component masses $m_1$ and $m_2$, where one of the masses may be negative and so we can define the total mass and reduced mass as
\begin{equation}
M = m_1 + m_2, \qquad \mu = \frac{m_1 m_2}{m_1 + m_2},
\end{equation}
the relative motion of the binary is governed by the equation
\begin{equation}
\ddot{\mathbf{r}} = -\frac{G M}{r^3} \mathbf{r},
\end{equation}
where $\mathbf{r}$ is the separation vector between the two bodies. As long as $M>0$, the system admits bound orbits analogous to Keplerian motion. But if one of the masses is negative, then the reduced mass $\mu$ becomes negative leading to qualitatively different energetic and radiative properties. The orbital energy of the system is given by
\begin{equation}
E = -\frac{G \mu M}{2a},
\end{equation}
where $a$ is the semi-major axis and for a conventional positive mass binary, $\mu>0$ and $E<0$, so the system is bound and loses energy through gravitational radiation. The quadrupole energy loss is given by
\begin{equation}
\dot{E}_{\rm GW} = -\frac{32}{5} \frac{G^4 \mu^2 M^3}{c^5 a^5},
\end{equation}
which is always negative, indicating that energy is radiated away. But when $\mu<0$ while $M>0$, the orbital energy becomes
\begin{equation}
E > 0,
\end{equation}
and the derivative of the energy with respect to separation is
\begin{equation}
\frac{dE}{da} = \frac{G \mu M}{2a^2} < 0.
\end{equation}
Combining this with the energy loss rate, one finds
\begin{equation}
\dot{a} = \frac{\dot{E}}{dE/da} > 0,
\end{equation}
so that the semi-major axis increases with time and so, instead of inspiraling, the binary expands while radiating energy. The gravitational wave frequency is related to the orbital frequency by
\begin{equation}
f_{\rm GW} = \frac{\Omega}{\pi}, \qquad \Omega^2 = \frac{G M}{a^3}
\end{equation}
Differentiating with respect to time gives
\begin{equation}
\dot{f}_{\rm GW} = \frac{df_{\rm GW}}{da} \dot{a}
\end{equation}
and using the scaling $\Omega \propto a^{-3/2}$, one finds
\begin{equation}
\dot{f}_{\rm GW} \propto -\dot{a}
\end{equation}
Since $\dot{a}>0$, it follows that
\begin{equation}
\dot{f}_{\rm GW} < 0.
\end{equation}
This corresponds to a decreasing gravitational wave frequency, in contrast to the increasing frequency observed in standard compact binary inspirals. More explicitly, the leading order frequency evolution can be written as \cite{lvk5berti2018extreme,lvk6carson2020probing,lvk7isi2017probing}
\begin{equation}
\dot{f}_{\rm GW} = \frac{96}{5} \pi^{8/3} \frac{G^{5/3}}{c^5} \mu M^{2/3} f_{\rm GW}^{11/3},
\end{equation}
so that the sign of $\dot{f}_{\rm GW}$ is determined by the sign of $\mu$. For $\mu>0$ one recovers the standard chirp signal with $\dot{f}_{\rm GW}>0$, while for $\mu<0$, one obtains
\begin{equation}
\dot{f}_{\rm GW} < 0,
\end{equation}
corresponding to an anti-chirp. This qualitative difference in frequency evolution leads to gravitational waveforms that are fundamentally distinct from those of positive mass binaries as in this case the quadrupole waveform itself,
\begin{equation}
h_{ij}^{TT} = \frac{2G}{c^4 D} \ddot{Q}_{ij}^{TT},
\end{equation}
with
\begin{equation}
Q_{ij} = \mu \left(r_i r_j - \frac{1}{3} \delta_{ij} r^2 \right),
\end{equation}
retains the same functional form, but the time evolution of the source is reversed in character. Instead of a chirp signal that increases in both amplitude and frequency, one expects a signal with decreasing frequency and a different amplitude evolution, reflecting the outward evolution of the binary.
\\
\\
An even more extreme case arises when the total mass vanishes, $M=0$, corresponding to $m_1 = -m_2$ and in this limit, the relative equation of motion reduces to
\begin{equation}
\ddot{\mathbf{r}} = 0
\end{equation}
so that the separation remains constant. At the same time, the individual bodies experience a common acceleration due to their mutual interaction leading to a runaway solution in which both objects accelerate indefinitely in the same direction. Such configurations do not correspond to bound binaries and therefore do not produce the periodic chirp signals characteristic of inspiraling systems. Instead, one expects highly non standard, non periodic gravitational wave emission, potentially including burst like or memory type signatures.
One can say that the absence of observed anti-chirp signals, as well as the lack of evidence for runaway or non-standard quadrupolar waveforms in gravitational wave data, places strong constraints on the existence of astrophysical systems containing negative masses. These constraints arise independently of any assumptions about additional gravitational degrees of freedom or modifications to the theory, and follow directly from the structure of the quadrupole radiation formula combined with the dynamics of negative mass systems. Note also that for positive mass/negative mass pair, Newton's force law can still be written in a 3rd law-type action-reaction form at the level of mutual forces, but the accelerations need not be opposite if one inertial mass is negative. This is also the origin of the Bondi type runaway behavior \cite{bondi1957negative}, wherein the forces may be equal and opposite while the negative inertial mass accelerates opposite to the applied force.
\\

To illustrate these constraints even more, let us consider three different cases of binaries. We first consider the case
\begin{equation}
m_1 = -A, \quad m_2 = +b, \quad A > b.
\end{equation}
Then
\begin{equation}
M = b - A < 0, \quad \mu = \frac{-Ab}{b-A} > 0.
\end{equation}
The equation of motion becomes
\begin{equation}
\ddot{\mathbf{r}} = +\frac{G |M|}{r^3} \mathbf{r},
\end{equation}
which corresponds to a repulsive interaction and no bound orbits exist in this case. The system disperses on a dynamical timescale. Assuming the objects start from rest at separation $r_0$, energy conservation gives
\begin{equation}
\frac{1}{2} \dot{r}^2 + \frac{G |M|}{r} = \frac{G |M|}{r_0}.
\end{equation}
Solving for $\dot{r}$
\begin{equation}
\dot{r} = \sqrt{2 G |M| \left( \frac{1}{r_0} - \frac{1}{r} \right)}.
\end{equation}
The expansion time to reach $r = \lambda r_0$ is
\begin{equation}
t(\lambda) =
\sqrt{\frac{r_0^3}{2 G |M|}}
\left[
\sqrt{\lambda(\lambda - 1)} +
\ln\left( \sqrt{\lambda} + \sqrt{\lambda - 1} \right)
\right].
\end{equation}
This shows that the system disperses on a timescale of order
\begin{equation}
t_{\rm dyn} \sim \sqrt{\frac{r_0^3}{G |M|}},
\end{equation}
which is much shorter than cosmological timescales and so no stable binary or gravitational wave chirp can arise. \\

For the second case, we now consider
\begin{equation}
m_1 = +A, \quad m_2 = -b, \quad A > b.
\end{equation}
Then
\begin{equation}
M = A - b > 0, \quad \mu = \frac{-Ab}{A-b} < 0.
\end{equation}
The relative motion allows circular orbits since $M>0$, but the orbital energy is
\begin{equation}
E = -\frac{G \mu M}{2a} > 0.
\end{equation}
Gravitational wave emission follows the quadrupole formula,
\begin{equation}
\dot{E}_{\rm GW} = -\frac{32}{5} \frac{G^4 \mu^2 M^3}{c^5 a^5},
\end{equation}
which is always negative but
\begin{equation}
\frac{dE}{da} = \frac{G \mu M}{2a^2} < 0,
\end{equation}
so the orbital evolution is
\begin{equation}
\dot{a} = \frac{\dot{E}}{dE/da} > 0.
\end{equation}
Thus the binary expands while radiating and the orbital frequency satisfies
\begin{equation}
\Omega^2 = \frac{G M}{a^3},
\end{equation}
so the gravitational wave frequency evolves as
\begin{equation}
\dot{f}_{\rm GW} =
\frac{96}{5} \pi^{8/3} \frac{G^{5/3}}{c^5} \mu M^{2/3} f_{\rm GW}^{11/3}.
\end{equation}
Since $\mu < 0$, one finds
\begin{equation}
\dot{f}_{\rm GW} < 0,
\end{equation}
corresponding to an anti-chirp with decreasing frequency. So the total radiated energy as the system expands to infinity is finite,
\begin{equation}
E_{\rm rad} = \frac{G |\mu| M}{2 a_0}.
\end{equation}
The characteristic timescale for expansion is
\begin{equation}
t_{\rm anti} \sim \frac{5}{64} \frac{c^5 a_0^4}{G^3 |\mu| M^2},
\end{equation}
which can be very long for astrophysical systems. For example, in the case that $a_0=r_0$, $|\mu|=10M_\odot$ and $M=20M_\odot$, one obtains $t_{\rm anti}\simeq 20\,{\rm s}\,(r_0/10^3\,{\rm km})^4$, implying that compact positive--negative systems in the gravitational wave band would produce short-lived anti-chirps rather than long lived inspirals. \\

\begin{figure}
\centering
\includegraphics[width=1.05\columnwidth]{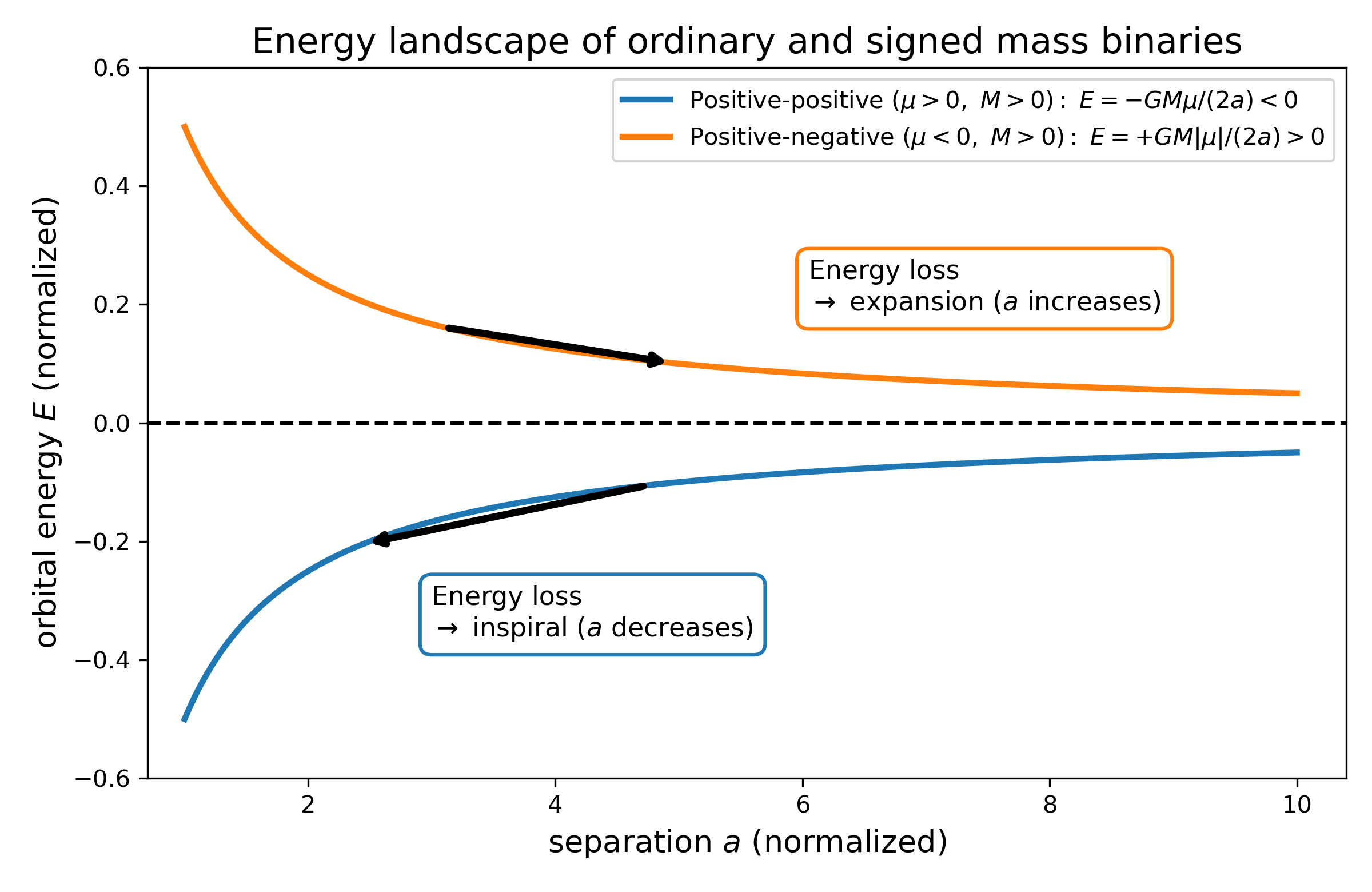}
\caption{Normalized orbital energy as a function of separation, showing that energy loss drives positive--positive binaries inward but positive-negative binaries outward.}
\label{signedmass}
\end{figure}

Finally, let us consider the case where
\begin{equation}
m_1 = -A, \quad m_2 = -B.
\end{equation}
Then
\begin{equation}
M = -(A+B) < 0, \quad \mu = \frac{AB}{-(A+B)} < 0.
\end{equation}
The equation of motion becomes
\begin{equation}
\ddot{\mathbf{r}} = +\frac{G (A+B)}{r^3} \mathbf{r},
\end{equation}
which is repulsive and so, no bound Keplerian orbits exist. Although we can see that the expression
\begin{equation}
E = -\frac{G \mu M}{2a}
\end{equation}
would formally give $E<0$, this expression assumes the existence of circular orbits which require $M>0$. Since this condition is not satisfied, the system cannot form a bound binary and instead disperses.
\\
\\
The above analysis shows that negative mass systems exhibit fundamentally different dynamical behavior from ordinary positive mass binaries. Systems with $M<0$ are repulsive and short lived, preventing the formation of stable binaries. Systems with $M>0$ but $\mu<0$ can form circular orbits, but their evolution under gravitational radiation leads to expansion rather than inspiral, producing anti-chirp signals with decreasing frequency. Systems with $M=0$ correspond to runaway solutions rather than periodic motion.
\begin{figure}
\centering
\includegraphics[width=1.05\columnwidth]{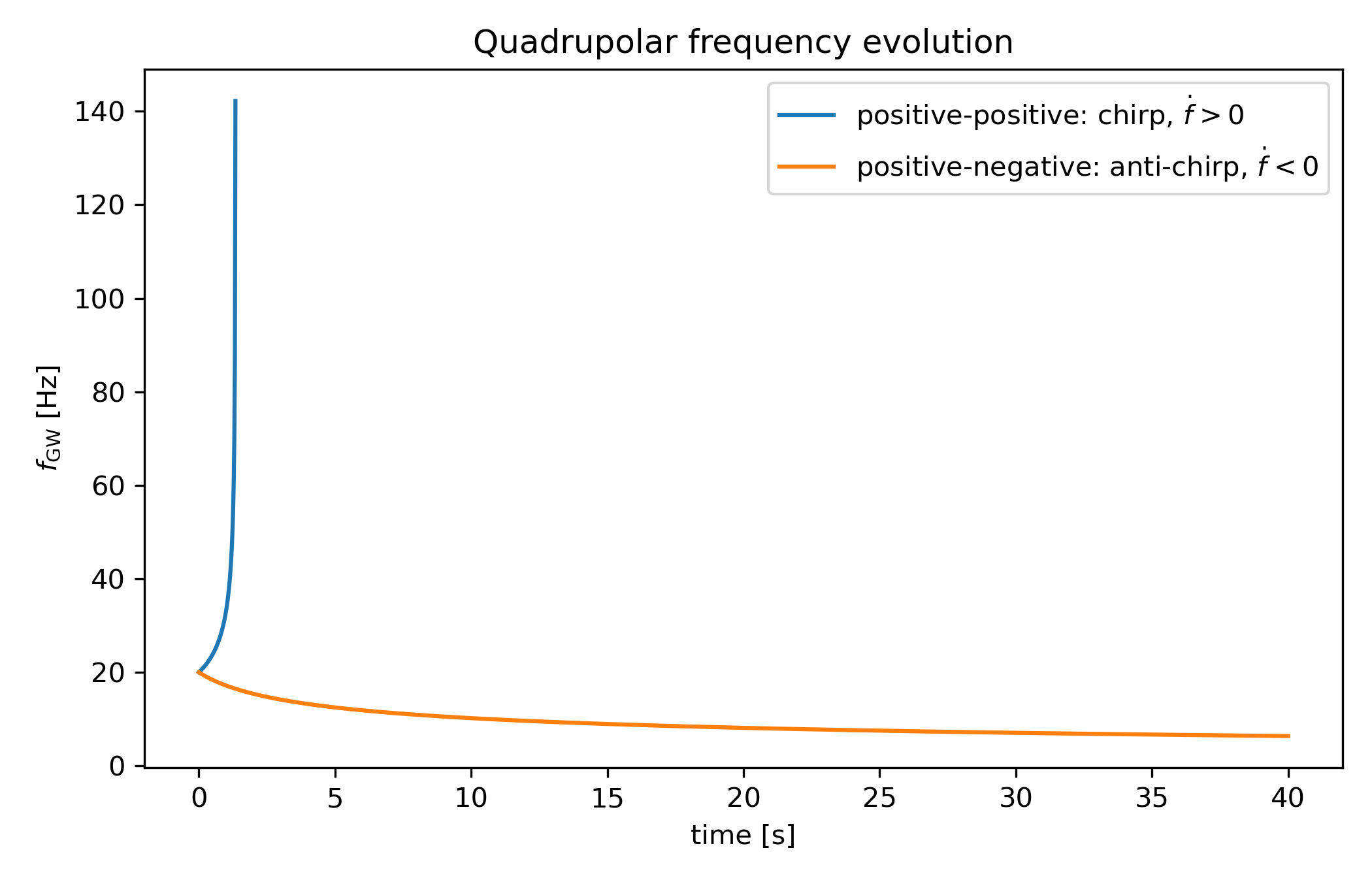}
\caption{Illustrative quadrupolar frequency evolution showing the standard chirp of a positive-positive binary and the anti-chirp produced by a positive-negative binary with $M>0$ and $\mu<0$.}
\label{distinct}
\end{figure}
We make the points raised here even more clear with figures \ref{distinct} and \ref{signedmass}. Figure \ref{distinct} illustrates the central waveform-level distinction between ordinary binaries and binaries containing a negative mass component. For a positive-positive system the reduced mass is positive, so the leading order quadrupole evolution gives $\dot f_{\rm GW}>0$, producing the familiar increasing frequency chirp observed by LIGO-Virgo-Kagra(LVK) \cite{lvk1abbott2019tests,lvk2abbott2021tests,lvk3ghosh2018testing}. But a positive-negative system with $M>0$ but $\mu<0$, the same quadrupole formula gives $\dot f_{\rm GW}<0$, so the system radiates while moving outward and its gravitational wave frequency decreases. The absence of such anti-chirp morphologies in current gravitational wave catalogs hence provides us with a direct observational handle on negative mass binaries \cite{lvk8wang2024tests,lvk9yunes2009fundamental,lvk10mishra2010parametrized}.
Figure \ref{signedmass} illustrates the energetic origin of the anti-chirp more visually. For an ordinary positive-positive binary, the orbital energy is negative and becomes more negative as the separation decreases, so gravitational wave energy loss drives the system toward inspiral and increasing frequency. For a positive-negative binary with $M>0$ and $\mu<0$, the orbital energy is instead positive and decreases toward zero as the separation grows, so the same loss of energy causes the system to expand rather than inspiral. \\ 

Throughout this work we distinguish the inertial mass $m_i$, which controls the response to a force, from the effective gravitational charge $q^X$, which controls the strength and sign of the gravitational coupling in a given channel. The coupling level constraints in the first part of the paper apply to the ratio $q^X/m_i$ and therefore test whether a negative-mass sector carries non-universal gravitational charge. The dynamical waveform analysis in the second part corresponds to the stronger signed-mass realization commonly assumed in negative mass discussions, in which the masses entering the two body problem including the reduced mass $\mu=m_1m_2/(m_1+m_2)$, may themselves be negative. Thus the dipole radiation constraints and the anti-chirp constraints probe related but distinct notions of negative mass, wherein the former constrains non-universal gravitational charge, while the latter constrains signed effective mass dynamics.

In conclusion, negative masses are viable only if they simultaneously satisfy universality of gravitational coupling across all observable channels and do not produce any dynamically induced signatures that are incompatible with observed gravitational wave phenomenology. Formally, this requires
\begin{equation}
\left(\frac{q^X}{m}\right)_- = \left(\frac{q^X}{m}\right)_+ \quad \text{for all observable channels } X,
\end{equation}
together with the requirement that their dynamical evolution does not generate waveform classes absent in current observations. Particularly the binary configurations involving negative masses must avoid regimes that generically produce anti-chirp behavior ($\dot f_{\rm GW}<0$), repulsive dispersal ($M<0$), or runaway solutions ($M=0$) and instead satisfy
\begin{equation}
h_{-}(t) \in \mathcal{H}_{\rm obs},
\end{equation}
where $\mathcal{H}_{\rm obs}$ denotes the set of gravitational wave morphologies consistent with current compact binary observations. Equivalently, the existence of any channel $X$ for which
\begin{equation}
\left|
\left(\frac{q^X}{m}\right)_+ - \left(\frac{q^X}{m}\right)_-
\right| > \epsilon_X,
\end{equation}
or any physically realized configuration leading to $\dot f_{\rm GW}<0$, $M<0$, or $M=0$ dynamics, implies exclusion of the corresponding negative mass model. \\

\textbf{Acknowledgements:}
The work of OT is supported in part by the Vanderbilt Discovery Doctoral Fellowship. The work of AL is supported in part by the Black Hole Initiative, which is funded by GBMF and JTF. 
\bibliography{negmprl}

\begin{thebibliography}{38}%
\makeatletter
\providecommand \@ifxundefined [1]{%
 \@ifx{#1\undefined}
}%
\providecommand \@ifnum [1]{%
 \ifnum #1\expandafter \@firstoftwo
 \else \expandafter \@secondoftwo
 \fi
}%
\providecommand \@ifx [1]{%
 \ifx #1\expandafter \@firstoftwo
 \else \expandafter \@secondoftwo
 \fi
}%
\providecommand \natexlab [1]{#1}%
\providecommand \enquote  [1]{``#1''}%
\providecommand \bibnamefont  [1]{#1}%
\providecommand \bibfnamefont [1]{#1}%
\providecommand \citenamefont [1]{#1}%
\providecommand \href@noop [0]{\@secondoftwo}%
\providecommand \href [0]{\begingroup \@sanitize@url \@href}%
\providecommand \@href[1]{\@@startlink{#1}\@@href}%
\providecommand \@@href[1]{\endgroup#1\@@endlink}%
\providecommand \@sanitize@url [0]{\catcode `\\12\catcode `\$12\catcode `\&12\catcode `\#12\catcode `\^12\catcode `\_12\catcode `\%12\relax}%
\providecommand \@@startlink[1]{}%
\providecommand \@@endlink[0]{}%
\providecommand \url  [0]{\begingroup\@sanitize@url \@url }%
\providecommand \@url [1]{\endgroup\@href {#1}{\urlprefix }}%
\providecommand \urlprefix  [0]{URL }%
\providecommand \Eprint [0]{\href }%
\providecommand \doibase [0]{https://doi.org/}%
\providecommand \selectlanguage [0]{\@gobble}%
\providecommand \bibinfo  [0]{\@secondoftwo}%
\providecommand \bibfield  [0]{\@secondoftwo}%
\providecommand \translation [1]{[#1]}%
\providecommand \BibitemOpen [0]{}%
\providecommand \bibitemStop [0]{}%
\providecommand \bibitemNoStop [0]{.\EOS\space}%
\providecommand \EOS [0]{\spacefactor3000\relax}%
\providecommand \BibitemShut  [1]{\csname bibitem#1\endcsname}%
\let\auto@bib@innerbib\@empty
\bibitem [{\citenamefont {Bondi}(1957)}]{bondi1957negative}%
  \BibitemOpen
  \bibfield  {author} {\bibinfo {author} {\bibfnamefont {H.}~\bibnamefont {Bondi}},\ }\bibfield  {title} {\bibinfo {title} {Negative mass in general relativity},\ }\href@noop {} {\bibfield  {journal} {\bibinfo  {journal} {Reviews of Modern Physics}\ }\textbf {\bibinfo {volume} {29}},\ \bibinfo {pages} {423} (\bibinfo {year} {1957})}\BibitemShut {NoStop}%
\bibitem [{\citenamefont {Manfredi}\ \emph {et~al.}(2026)\citenamefont {Manfredi}, \citenamefont {Rouet},\ and\ \citenamefont {Miller}}]{nm1manfredi2026bondi}%
  \BibitemOpen
  \bibfield  {author} {\bibinfo {author} {\bibfnamefont {G.}~\bibnamefont {Manfredi}}, \bibinfo {author} {\bibfnamefont {J.-L.}\ \bibnamefont {Rouet}},\ and\ \bibinfo {author} {\bibfnamefont {B.}~\bibnamefont {Miller}},\ }\bibfield  {title} {\bibinfo {title} {The bondi universe: How negative mass drives the cosmological expansion},\ }\href@noop {} {\bibfield  {journal} {\bibinfo  {journal} {arXiv preprint arXiv:2601.22910}\ } (\bibinfo {year} {2026})}\BibitemShut {NoStop}%
\bibitem [{\citenamefont {Nojiri}\ and\ \citenamefont {Odintsov}(2026)}]{nm2nojiri2026may}%
  \BibitemOpen
  \bibfield  {author} {\bibinfo {author} {\bibfnamefont {S.}~\bibnamefont {Nojiri}}\ and\ \bibinfo {author} {\bibfnamefont {S.}~\bibnamefont {Odintsov}},\ }\bibfield  {title} {\bibinfo {title} {May negative mass objects exist in the sky?},\ }\href@noop {} {\bibfield  {journal} {\bibinfo  {journal} {arXiv preprint arXiv:2602.15058}\ } (\bibinfo {year} {2026})}\BibitemShut {NoStop}%
\bibitem [{\citenamefont {Farnes}(2018)}]{nm3farnes2018unifying}%
  \BibitemOpen
  \bibfield  {author} {\bibinfo {author} {\bibfnamefont {J.~S.}\ \bibnamefont {Farnes}},\ }\bibfield  {title} {\bibinfo {title} {A unifying theory of dark energy and dark matter: Negative masses and matter creation within a modified $\lambda$cdm framework},\ }\href@noop {} {\bibfield  {journal} {\bibinfo  {journal} {Astronomy \& Astrophysics}\ }\textbf {\bibinfo {volume} {620}},\ \bibinfo {pages} {A92} (\bibinfo {year} {2018})}\BibitemShut {NoStop}%
\bibitem [{\citenamefont {Bellet{\^e}te}\ and\ \citenamefont {Paranjape}(2013)}]{nm4belletete2013negative}%
  \BibitemOpen
  \bibfield  {author} {\bibinfo {author} {\bibfnamefont {J.}~\bibnamefont {Bellet{\^e}te}}\ and\ \bibinfo {author} {\bibfnamefont {M.}~\bibnamefont {Paranjape}},\ }\bibfield  {title} {\bibinfo {title} {On negative mass},\ }\href@noop {} {\bibfield  {journal} {\bibinfo  {journal} {International Journal of Modern Physics D}\ }\textbf {\bibinfo {volume} {22}},\ \bibinfo {pages} {1341017} (\bibinfo {year} {2013})}\BibitemShut {NoStop}%
\bibitem [{\citenamefont {Bonnor}(1989)}]{nm5bonnor1989negative}%
  \BibitemOpen
  \bibfield  {author} {\bibinfo {author} {\bibfnamefont {W.~B.}\ \bibnamefont {Bonnor}},\ }\bibfield  {title} {\bibinfo {title} {Negative mass in general relativity},\ }\href@noop {} {\bibfield  {journal} {\bibinfo  {journal} {General relativity and gravitation}\ }\textbf {\bibinfo {volume} {21}},\ \bibinfo {pages} {1143} (\bibinfo {year} {1989})}\BibitemShut {NoStop}%
\bibitem [{\citenamefont {Cebeci}\ \emph {et~al.}(2006)\citenamefont {Cebeci}, \citenamefont {Sar{\i}o{\u{g}}lu},\ and\ \citenamefont {Tekin}}]{nm13cebeci2006negative}%
  \BibitemOpen
  \bibfield  {author} {\bibinfo {author} {\bibfnamefont {H.}~\bibnamefont {Cebeci}}, \bibinfo {author} {\bibfnamefont {{\"O}.}~\bibnamefont {Sar{\i}o{\u{g}}lu}},\ and\ \bibinfo {author} {\bibfnamefont {B.}~\bibnamefont {Tekin}},\ }\bibfield  {title} {\bibinfo {title} {Negative mass solitons in gravity},\ }\href@noop {} {\bibfield  {journal} {\bibinfo  {journal} {Physical Review D—Particles, Fields, Gravitation, and Cosmology}\ }\textbf {\bibinfo {volume} {73}},\ \bibinfo {pages} {064020} (\bibinfo {year} {2006})}\BibitemShut {NoStop}%
\bibitem [{\citenamefont {Mann}(1997)}]{nm14mann1997black}%
  \BibitemOpen
  \bibfield  {author} {\bibinfo {author} {\bibfnamefont {R.}~\bibnamefont {Mann}},\ }\bibfield  {title} {\bibinfo {title} {Black holes of negative mass},\ }\href@noop {} {\bibfield  {journal} {\bibinfo  {journal} {Classical and Quantum Gravity}\ }\textbf {\bibinfo {volume} {14}},\ \bibinfo {pages} {2927} (\bibinfo {year} {1997})}\BibitemShut {NoStop}%
\bibitem [{\citenamefont {Forward}(1990)}]{forward1990negative}%
  \BibitemOpen
  \bibfield  {author} {\bibinfo {author} {\bibfnamefont {R.~L.}\ \bibnamefont {Forward}},\ }\bibfield  {title} {\bibinfo {title} {Negative matter propulsion},\ }\href@noop {} {\bibfield  {journal} {\bibinfo  {journal} {Journal of Propulsion and Power}\ }\textbf {\bibinfo {volume} {6}},\ \bibinfo {pages} {28} (\bibinfo {year} {1990})}\BibitemShut {NoStop}%
\bibitem [{\citenamefont {M{\o}ller}\ \emph {et~al.}(2017)\citenamefont {M{\o}ller}, \citenamefont {Thomas}, \citenamefont {Vasilakis}, \citenamefont {Zeuthen}, \citenamefont {Tsaturyan}, \citenamefont {Balabas}, \citenamefont {Jensen}, \citenamefont {Schliesser}, \citenamefont {Hammerer},\ and\ \citenamefont {Polzik}}]{nm6moller2017quantum}%
  \BibitemOpen
  \bibfield  {author} {\bibinfo {author} {\bibfnamefont {C.~B.}\ \bibnamefont {M{\o}ller}}, \bibinfo {author} {\bibfnamefont {R.~A.}\ \bibnamefont {Thomas}}, \bibinfo {author} {\bibfnamefont {G.}~\bibnamefont {Vasilakis}}, \bibinfo {author} {\bibfnamefont {E.}~\bibnamefont {Zeuthen}}, \bibinfo {author} {\bibfnamefont {Y.}~\bibnamefont {Tsaturyan}}, \bibinfo {author} {\bibfnamefont {M.}~\bibnamefont {Balabas}}, \bibinfo {author} {\bibfnamefont {K.}~\bibnamefont {Jensen}}, \bibinfo {author} {\bibfnamefont {A.}~\bibnamefont {Schliesser}}, \bibinfo {author} {\bibfnamefont {K.}~\bibnamefont {Hammerer}},\ and\ \bibinfo {author} {\bibfnamefont {E.~S.}\ \bibnamefont {Polzik}},\ }\bibfield  {title} {\bibinfo {title} {Quantum back-action-evading measurement of motion in a negative mass reference frame},\ }\href@noop {} {\bibfield  {journal} {\bibinfo  {journal} {Nature}\ }\textbf {\bibinfo {volume} {547}},\ \bibinfo {pages} {191} (\bibinfo {year} {2017})}\BibitemShut {NoStop}%
\bibitem [{\citenamefont {Khamehchi}\ \emph {et~al.}(2017)\citenamefont {Khamehchi}, \citenamefont {Hossain}, \citenamefont {Mossman}, \citenamefont {Zhang}, \citenamefont {Busch}, \citenamefont {Forbes},\ and\ \citenamefont {Engels}}]{nm7khamehchi2017negative}%
  \BibitemOpen
  \bibfield  {author} {\bibinfo {author} {\bibfnamefont {M.}~\bibnamefont {Khamehchi}}, \bibinfo {author} {\bibfnamefont {K.}~\bibnamefont {Hossain}}, \bibinfo {author} {\bibfnamefont {M.}~\bibnamefont {Mossman}}, \bibinfo {author} {\bibfnamefont {Y.}~\bibnamefont {Zhang}}, \bibinfo {author} {\bibfnamefont {T.}~\bibnamefont {Busch}}, \bibinfo {author} {\bibfnamefont {M.~M.}\ \bibnamefont {Forbes}},\ and\ \bibinfo {author} {\bibfnamefont {P.}~\bibnamefont {Engels}},\ }\bibfield  {title} {\bibinfo {title} {Negative-mass hydrodynamics in a spin-orbit--coupled bose-einstein condensate},\ }\href@noop {} {\bibfield  {journal} {\bibinfo  {journal} {Physical review letters}\ }\textbf {\bibinfo {volume} {118}},\ \bibinfo {pages} {155301} (\bibinfo {year} {2017})}\BibitemShut {NoStop}%
\bibitem [{\citenamefont {Socas-Navarro}(2019)}]{nm9socas2019can}%
  \BibitemOpen
  \bibfield  {author} {\bibinfo {author} {\bibfnamefont {H.}~\bibnamefont {Socas-Navarro}},\ }\bibfield  {title} {\bibinfo {title} {Can a negative-mass cosmology explain dark matter and dark energy?},\ }\href@noop {} {\bibfield  {journal} {\bibinfo  {journal} {Astronomy \& Astrophysics}\ }\textbf {\bibinfo {volume} {626}},\ \bibinfo {pages} {A5} (\bibinfo {year} {2019})}\BibitemShut {NoStop}%
\bibitem [{\citenamefont {Petit}\ and\ \citenamefont {d’Agostini}(2014)}]{nm10petit2014negative}%
  \BibitemOpen
  \bibfield  {author} {\bibinfo {author} {\bibfnamefont {J.-P.}\ \bibnamefont {Petit}}\ and\ \bibinfo {author} {\bibfnamefont {G.}~\bibnamefont {d’Agostini}},\ }\bibfield  {title} {\bibinfo {title} {Negative mass hypothesis in cosmology and the nature of dark energy},\ }\href@noop {} {\bibfield  {journal} {\bibinfo  {journal} {Astrophysics and Space Science}\ }\textbf {\bibinfo {volume} {354}},\ \bibinfo {pages} {611} (\bibinfo {year} {2014})}\BibitemShut {NoStop}%
\bibitem [{\citenamefont {N{\'a}jera}\ \emph {et~al.}(2021)\citenamefont {N{\'a}jera}, \citenamefont {Gamboa}, \citenamefont {Aguilar-Nieto},\ and\ \citenamefont {Escamilla-Rivera}}]{nm11najera2021negative}%
  \BibitemOpen
  \bibfield  {author} {\bibinfo {author} {\bibfnamefont {S.}~\bibnamefont {N{\'a}jera}}, \bibinfo {author} {\bibfnamefont {A.}~\bibnamefont {Gamboa}}, \bibinfo {author} {\bibfnamefont {A.}~\bibnamefont {Aguilar-Nieto}},\ and\ \bibinfo {author} {\bibfnamefont {C.}~\bibnamefont {Escamilla-Rivera}},\ }\bibfield  {title} {\bibinfo {title} {On negative mass cosmology in general relativity},\ }\href@noop {} {\bibfield  {journal} {\bibinfo  {journal} {Astronomy \& Astrophysics}\ }\textbf {\bibinfo {volume} {651}},\ \bibinfo {pages} {L13} (\bibinfo {year} {2021})}\BibitemShut {NoStop}%
\bibitem [{\citenamefont {Manfredi}\ \emph {et~al.}(2018)\citenamefont {Manfredi}, \citenamefont {Rouet}, \citenamefont {Miller},\ and\ \citenamefont {Chardin}}]{nm12manfredi2018cosmological}%
  \BibitemOpen
  \bibfield  {author} {\bibinfo {author} {\bibfnamefont {G.}~\bibnamefont {Manfredi}}, \bibinfo {author} {\bibfnamefont {J.-L.}\ \bibnamefont {Rouet}}, \bibinfo {author} {\bibfnamefont {B.}~\bibnamefont {Miller}},\ and\ \bibinfo {author} {\bibfnamefont {G.}~\bibnamefont {Chardin}},\ }\bibfield  {title} {\bibinfo {title} {Cosmological structure formation with negative mass},\ }\href@noop {} {\bibfield  {journal} {\bibinfo  {journal} {Physical Review D}\ }\textbf {\bibinfo {volume} {98}},\ \bibinfo {pages} {023514} (\bibinfo {year} {2018})}\BibitemShut {NoStop}%
\bibitem [{\citenamefont {Wang}\ \emph {et~al.}(2022)\citenamefont {Wang}, \citenamefont {Zhao}, \citenamefont {An}, \citenamefont {Shao},\ and\ \citenamefont {Cao}}]{dip1boundwang2022simultaneous}%
  \BibitemOpen
  \bibfield  {author} {\bibinfo {author} {\bibfnamefont {Z.}~\bibnamefont {Wang}}, \bibinfo {author} {\bibfnamefont {J.}~\bibnamefont {Zhao}}, \bibinfo {author} {\bibfnamefont {Z.}~\bibnamefont {An}}, \bibinfo {author} {\bibfnamefont {L.}~\bibnamefont {Shao}},\ and\ \bibinfo {author} {\bibfnamefont {Z.}~\bibnamefont {Cao}},\ }\bibfield  {title} {\bibinfo {title} {Simultaneous bounds on the gravitational dipole radiation and varying gravitational constant from compact binary inspirals},\ }\href@noop {} {\bibfield  {journal} {\bibinfo  {journal} {Physics Letters B}\ }\textbf {\bibinfo {volume} {834}},\ \bibinfo {pages} {137416} (\bibinfo {year} {2022})}\BibitemShut {NoStop}%
\bibitem [{\citenamefont {Lazaridis}\ \emph {et~al.}(2009)\citenamefont {Lazaridis}, \citenamefont {Wex}, \citenamefont {Jessner}, \citenamefont {Kramer}, \citenamefont {Stappers}, \citenamefont {Janssen}, \citenamefont {Desvignes}, \citenamefont {Purver}, \citenamefont {Cognard}, \citenamefont {Theureau} \emph {et~al.}}]{dip2lazaridis2009generic}%
  \BibitemOpen
  \bibfield  {author} {\bibinfo {author} {\bibfnamefont {K.}~\bibnamefont {Lazaridis}}, \bibinfo {author} {\bibfnamefont {N.}~\bibnamefont {Wex}}, \bibinfo {author} {\bibfnamefont {A.}~\bibnamefont {Jessner}}, \bibinfo {author} {\bibfnamefont {M.}~\bibnamefont {Kramer}}, \bibinfo {author} {\bibfnamefont {B.}~\bibnamefont {Stappers}}, \bibinfo {author} {\bibfnamefont {G.}~\bibnamefont {Janssen}}, \bibinfo {author} {\bibfnamefont {G.}~\bibnamefont {Desvignes}}, \bibinfo {author} {\bibfnamefont {M.}~\bibnamefont {Purver}}, \bibinfo {author} {\bibfnamefont {I.}~\bibnamefont {Cognard}}, \bibinfo {author} {\bibfnamefont {G.}~\bibnamefont {Theureau}}, \emph {et~al.},\ }\bibfield  {title} {\bibinfo {title} {Generic tests of the existence of the gravitational dipole radiation and the variation of the gravitational constant},\ }\href@noop {} {\bibfield  {journal} {\bibinfo  {journal} {Monthly Notices of the Royal Astronomical Society}\ }\textbf {\bibinfo {volume} {400}},\ \bibinfo {pages} {805} (\bibinfo {year}
  {2009})}\BibitemShut {NoStop}%
\bibitem [{\citenamefont {Will}(1977)}]{dip3will1977gravitational}%
  \BibitemOpen
  \bibfield  {author} {\bibinfo {author} {\bibfnamefont {C.~M.}\ \bibnamefont {Will}},\ }\bibfield  {title} {\bibinfo {title} {Gravitational radiation from binary systems in alternative metric theories of gravity-dipole radiation and the binary pulsar},\ }\href@noop {} {\bibfield  {journal} {\bibinfo  {journal} {Astrophysical Journal, Part 1, vol. 214, June 15, 1977, p. 826-839.}\ }\textbf {\bibinfo {volume} {214}},\ \bibinfo {pages} {826} (\bibinfo {year} {1977})}\BibitemShut {NoStop}%
\bibitem [{\citenamefont {Krisher}(1985)}]{dip4krisher1985dipole}%
  \BibitemOpen
  \bibfield  {author} {\bibinfo {author} {\bibfnamefont {T.~P.}\ \bibnamefont {Krisher}},\ }\bibfield  {title} {\bibinfo {title} {Dipole gravitational radiation in the nonsymmetric gravitational theory of moffat},\ }\href@noop {} {\bibfield  {journal} {\bibinfo  {journal} {Physical Review D}\ }\textbf {\bibinfo {volume} {32}},\ \bibinfo {pages} {329} (\bibinfo {year} {1985})}\BibitemShut {NoStop}%
\bibitem [{\citenamefont {Barausse}\ \emph {et~al.}(2016)\citenamefont {Barausse}, \citenamefont {Yunes},\ and\ \citenamefont {Chamberlain}}]{dip5barausse2016theory}%
  \BibitemOpen
  \bibfield  {author} {\bibinfo {author} {\bibfnamefont {E.}~\bibnamefont {Barausse}}, \bibinfo {author} {\bibfnamefont {N.}~\bibnamefont {Yunes}},\ and\ \bibinfo {author} {\bibfnamefont {K.}~\bibnamefont {Chamberlain}},\ }\bibfield  {title} {\bibinfo {title} {Theory-agnostic constraints on black-hole dipole radiation with multiband gravitational-wave astrophysics},\ }\href@noop {} {\bibfield  {journal} {\bibinfo  {journal} {Physical review letters}\ }\textbf {\bibinfo {volume} {116}},\ \bibinfo {pages} {241104} (\bibinfo {year} {2016})}\BibitemShut {NoStop}%
\bibitem [{\citenamefont {Vas{\'u}th}\ \emph {et~al.}(2003)\citenamefont {Vas{\'u}th}, \citenamefont {Keresztes}, \citenamefont {Mih{\'a}ly},\ and\ \citenamefont {Gergely}}]{dip6vasuth2003gravitational}%
  \BibitemOpen
  \bibfield  {author} {\bibinfo {author} {\bibfnamefont {M.}~\bibnamefont {Vas{\'u}th}}, \bibinfo {author} {\bibfnamefont {Z.}~\bibnamefont {Keresztes}}, \bibinfo {author} {\bibfnamefont {A.}~\bibnamefont {Mih{\'a}ly}},\ and\ \bibinfo {author} {\bibfnamefont {L.~{\'A}.}\ \bibnamefont {Gergely}},\ }\bibfield  {title} {\bibinfo {title} {Gravitational radiation reaction in compact binary systems: Contribution of the magnetic dipole--magnetic dipole interaction},\ }\href@noop {} {\bibfield  {journal} {\bibinfo  {journal} {Physical Review D}\ }\textbf {\bibinfo {volume} {68}},\ \bibinfo {pages} {124006} (\bibinfo {year} {2003})}\BibitemShut {NoStop}%
\bibitem [{\citenamefont {Weisberg}\ and\ \citenamefont {Taylor}(1981)}]{dip7weisberg1981gravitational}%
  \BibitemOpen
  \bibfield  {author} {\bibinfo {author} {\bibfnamefont {J.~M.}\ \bibnamefont {Weisberg}}\ and\ \bibinfo {author} {\bibfnamefont {J.~H.}\ \bibnamefont {Taylor}},\ }\bibfield  {title} {\bibinfo {title} {Gravitational radiation from an orbiting pulsar},\ }\href@noop {} {\bibfield  {journal} {\bibinfo  {journal} {General Relativity and Gravitation}\ }\textbf {\bibinfo {volume} {13}},\ \bibinfo {pages} {1} (\bibinfo {year} {1981})}\BibitemShut {NoStop}%
\bibitem [{\citenamefont {Arun}(2012)}]{dip8arun2012generic}%
  \BibitemOpen
  \bibfield  {author} {\bibinfo {author} {\bibfnamefont {K.}~\bibnamefont {Arun}},\ }\bibfield  {title} {\bibinfo {title} {Generic bounds on dipolar gravitational radiation from inspiralling compact binaries},\ }\href@noop {} {\bibfield  {journal} {\bibinfo  {journal} {Classical and Quantum Gravity}\ }\textbf {\bibinfo {volume} {29}},\ \bibinfo {pages} {075011} (\bibinfo {year} {2012})}\BibitemShut {NoStop}%
\bibitem [{\citenamefont {Yagi}\ \emph {et~al.}(2016)\citenamefont {Yagi}, \citenamefont {Stein},\ and\ \citenamefont {Yunes}}]{fut1yagi2016challenging}%
  \BibitemOpen
  \bibfield  {author} {\bibinfo {author} {\bibfnamefont {K.}~\bibnamefont {Yagi}}, \bibinfo {author} {\bibfnamefont {L.~C.}\ \bibnamefont {Stein}},\ and\ \bibinfo {author} {\bibfnamefont {N.}~\bibnamefont {Yunes}},\ }\bibfield  {title} {\bibinfo {title} {Challenging the presence of scalar charge and dipolar radiation in binary pulsars},\ }\href@noop {} {\bibfield  {journal} {\bibinfo  {journal} {Physical Review D}\ }\textbf {\bibinfo {volume} {93}},\ \bibinfo {pages} {024010} (\bibinfo {year} {2016})}\BibitemShut {NoStop}%
\bibitem [{\citenamefont {Barausse}\ and\ \citenamefont {Yagi}(2015)}]{fut2barausse2015gravitation}%
  \BibitemOpen
  \bibfield  {author} {\bibinfo {author} {\bibfnamefont {E.}~\bibnamefont {Barausse}}\ and\ \bibinfo {author} {\bibfnamefont {K.}~\bibnamefont {Yagi}},\ }\bibfield  {title} {\bibinfo {title} {Gravitation-wave emission in shift-symmetric horndeski theories},\ }\href@noop {} {\bibfield  {journal} {\bibinfo  {journal} {Physical Review Letters}\ }\textbf {\bibinfo {volume} {115}},\ \bibinfo {pages} {211105} (\bibinfo {year} {2015})}\BibitemShut {NoStop}%
\bibitem [{\citenamefont {Barausse}(2017)}]{fut3barausse2017testing}%
  \BibitemOpen
  \bibfield  {author} {\bibinfo {author} {\bibfnamefont {E.}~\bibnamefont {Barausse}},\ }\bibfield  {title} {\bibinfo {title} {Testing the strong equivalence principle with gravitational-wave observations of binary black holes},\ }\href@noop {} {\bibfield  {journal} {\bibinfo  {journal} {arXiv preprint arXiv:1703.05699}\ } (\bibinfo {year} {2017})}\BibitemShut {NoStop}%
\bibitem [{\citenamefont {Das}\ \emph {et~al.}(2025)\citenamefont {Das}, \citenamefont {Fridman},\ and\ \citenamefont {Lambiase}}]{fut4das2025testing}%
  \BibitemOpen
  \bibfield  {author} {\bibinfo {author} {\bibfnamefont {S.}~\bibnamefont {Das}}, \bibinfo {author} {\bibfnamefont {M.}~\bibnamefont {Fridman}},\ and\ \bibinfo {author} {\bibfnamefont {G.}~\bibnamefont {Lambiase}},\ }\bibfield  {title} {\bibinfo {title} {Testing the quantum equivalence principle with gravitational waves},\ }\href@noop {} {\bibfield  {journal} {\bibinfo  {journal} {Journal of High Energy Astrophysics}\ }\textbf {\bibinfo {volume} {48}},\ \bibinfo {pages} {100413} (\bibinfo {year} {2025})}\BibitemShut {NoStop}%
\bibitem [{\citenamefont {Lambiase}\ \emph {et~al.}(2021)\citenamefont {Lambiase}, \citenamefont {Sakellariadou},\ and\ \citenamefont {Stabile}}]{fut5lambiase2021constraints}%
  \BibitemOpen
  \bibfield  {author} {\bibinfo {author} {\bibfnamefont {G.}~\bibnamefont {Lambiase}}, \bibinfo {author} {\bibfnamefont {M.}~\bibnamefont {Sakellariadou}},\ and\ \bibinfo {author} {\bibfnamefont {A.}~\bibnamefont {Stabile}},\ }\bibfield  {title} {\bibinfo {title} {Constraints on extended gravity models through gravitational wave emission},\ }\href@noop {} {\bibfield  {journal} {\bibinfo  {journal} {Journal of Cosmology and Astroparticle Physics}\ }\textbf {\bibinfo {volume} {2021}}\bibinfo  {number} { (03)},\ \bibinfo {pages} {014}}\BibitemShut {NoStop}%
\bibitem [{\citenamefont {Abbott}\ \emph {et~al.}(2019)\citenamefont {Abbott}, \citenamefont {Abbott}, \citenamefont {Abbott}, \citenamefont {Abraham}, \citenamefont {Acernese}, \citenamefont {Ackley}, \citenamefont {Adams}, \citenamefont {Adhikari}, \citenamefont {Adya}, \citenamefont {Affeldt} \emph {et~al.}}]{lvk1abbott2019tests}%
  \BibitemOpen
\bibfield  {number} {  }\bibfield  {author} {\bibinfo {author} {\bibfnamefont {B.}~\bibnamefont {Abbott}}, \bibinfo {author} {\bibfnamefont {R.}~\bibnamefont {Abbott}}, \bibinfo {author} {\bibfnamefont {T.}~\bibnamefont {Abbott}}, \bibinfo {author} {\bibfnamefont {S.}~\bibnamefont {Abraham}}, \bibinfo {author} {\bibfnamefont {F.}~\bibnamefont {Acernese}}, \bibinfo {author} {\bibfnamefont {K.}~\bibnamefont {Ackley}}, \bibinfo {author} {\bibfnamefont {C.}~\bibnamefont {Adams}}, \bibinfo {author} {\bibfnamefont {R.~X.}\ \bibnamefont {Adhikari}}, \bibinfo {author} {\bibfnamefont {V.}~\bibnamefont {Adya}}, \bibinfo {author} {\bibfnamefont {C.}~\bibnamefont {Affeldt}}, \emph {et~al.},\ }\bibfield  {title} {\bibinfo {title} {Tests of general relativity with the binary black hole signals from the ligo-virgo catalog gwtc-1},\ }\href@noop {} {\bibfield  {journal} {\bibinfo  {journal} {Physical Review D}\ }\textbf {\bibinfo {volume} {100}},\ \bibinfo {pages} {104036} (\bibinfo {year} {2019})}\BibitemShut {NoStop}%
\bibitem [{\citenamefont {Abbott}\ \emph {et~al.}(2021)\citenamefont {Abbott}, \citenamefont {Abbott}, \citenamefont {Abraham}, \citenamefont {Acernese}, \citenamefont {Ackley}, \citenamefont {Adams}, \citenamefont {Adams}, \citenamefont {Adhikari}, \citenamefont {Adya}, \citenamefont {Affeldt} \emph {et~al.}}]{lvk2abbott2021tests}%
  \BibitemOpen
  \bibfield  {author} {\bibinfo {author} {\bibfnamefont {R.}~\bibnamefont {Abbott}}, \bibinfo {author} {\bibfnamefont {T.}~\bibnamefont {Abbott}}, \bibinfo {author} {\bibfnamefont {S.}~\bibnamefont {Abraham}}, \bibinfo {author} {\bibfnamefont {F.}~\bibnamefont {Acernese}}, \bibinfo {author} {\bibfnamefont {K.}~\bibnamefont {Ackley}}, \bibinfo {author} {\bibfnamefont {A.}~\bibnamefont {Adams}}, \bibinfo {author} {\bibfnamefont {C.}~\bibnamefont {Adams}}, \bibinfo {author} {\bibfnamefont {R.~X.}\ \bibnamefont {Adhikari}}, \bibinfo {author} {\bibfnamefont {V.}~\bibnamefont {Adya}}, \bibinfo {author} {\bibfnamefont {C.}~\bibnamefont {Affeldt}}, \emph {et~al.},\ }\bibfield  {title} {\bibinfo {title} {Tests of general relativity with binary black holes from the second ligo-virgo gravitational-wave transient catalog},\ }\href@noop {} {\bibfield  {journal} {\bibinfo  {journal} {Physical review D}\ }\textbf {\bibinfo {volume} {103}},\ \bibinfo {pages} {122002} (\bibinfo {year} {2021})}\BibitemShut {NoStop}%
\bibitem [{\citenamefont {Ghosh}\ \emph {et~al.}(2018)\citenamefont {Ghosh}, \citenamefont {Johnson-McDaniel}, \citenamefont {Ghosh}, \citenamefont {Mishra}, \citenamefont {Ajith}, \citenamefont {Pozzo}, \citenamefont {Berry}, \citenamefont {Nielsen},\ and\ \citenamefont {London}}]{lvk3ghosh2018testing}%
  \BibitemOpen
  \bibfield  {author} {\bibinfo {author} {\bibfnamefont {A.}~\bibnamefont {Ghosh}}, \bibinfo {author} {\bibfnamefont {N.~K.}\ \bibnamefont {Johnson-McDaniel}}, \bibinfo {author} {\bibfnamefont {A.}~\bibnamefont {Ghosh}}, \bibinfo {author} {\bibfnamefont {C.~K.}\ \bibnamefont {Mishra}}, \bibinfo {author} {\bibfnamefont {P.}~\bibnamefont {Ajith}}, \bibinfo {author} {\bibfnamefont {W.~D.}\ \bibnamefont {Pozzo}}, \bibinfo {author} {\bibfnamefont {C.~P.}\ \bibnamefont {Berry}}, \bibinfo {author} {\bibfnamefont {A.~B.}\ \bibnamefont {Nielsen}},\ and\ \bibinfo {author} {\bibfnamefont {L.}~\bibnamefont {London}},\ }\bibfield  {title} {\bibinfo {title} {Testing general relativity using gravitational wave signals from the inspiral, merger and ringdown of binary black holes},\ }\href@noop {} {\bibfield  {journal} {\bibinfo  {journal} {Classical and Quantum Gravity}\ }\textbf {\bibinfo {volume} {35}},\ \bibinfo {pages} {014002} (\bibinfo {year} {2018})}\BibitemShut {NoStop}%
\bibitem [{\citenamefont {Yunes}\ \emph {et~al.}(2016)\citenamefont {Yunes}, \citenamefont {Yagi},\ and\ \citenamefont {Pretorius}}]{lvk4yunes2016theoretical}%
  \BibitemOpen
  \bibfield  {author} {\bibinfo {author} {\bibfnamefont {N.}~\bibnamefont {Yunes}}, \bibinfo {author} {\bibfnamefont {K.}~\bibnamefont {Yagi}},\ and\ \bibinfo {author} {\bibfnamefont {F.}~\bibnamefont {Pretorius}},\ }\bibfield  {title} {\bibinfo {title} {Theoretical physics implications of the binary black-hole mergers gw150914 and gw151226},\ }\href@noop {} {\bibfield  {journal} {\bibinfo  {journal} {Physical review D}\ }\textbf {\bibinfo {volume} {94}},\ \bibinfo {pages} {084002} (\bibinfo {year} {2016})}\BibitemShut {NoStop}%
\bibitem [{\citenamefont {Berti}\ \emph {et~al.}(2018)\citenamefont {Berti}, \citenamefont {Yagi},\ and\ \citenamefont {Yunes}}]{lvk5berti2018extreme}%
  \BibitemOpen
  \bibfield  {author} {\bibinfo {author} {\bibfnamefont {E.}~\bibnamefont {Berti}}, \bibinfo {author} {\bibfnamefont {K.}~\bibnamefont {Yagi}},\ and\ \bibinfo {author} {\bibfnamefont {N.}~\bibnamefont {Yunes}},\ }\bibfield  {title} {\bibinfo {title} {Extreme gravity tests with gravitational waves from compact binary coalescences:(i) inspiral--merger},\ }\href@noop {} {\bibfield  {journal} {\bibinfo  {journal} {General Relativity and Gravitation}\ }\textbf {\bibinfo {volume} {50}},\ \bibinfo {pages} {46} (\bibinfo {year} {2018})}\BibitemShut {NoStop}%
\bibitem [{\citenamefont {Carson}\ and\ \citenamefont {Yagi}(2020)}]{lvk6carson2020probing}%
  \BibitemOpen
  \bibfield  {author} {\bibinfo {author} {\bibfnamefont {Z.}~\bibnamefont {Carson}}\ and\ \bibinfo {author} {\bibfnamefont {K.}~\bibnamefont {Yagi}},\ }\bibfield  {title} {\bibinfo {title} {Probing beyond-kerr spacetimes with inspiral-ringdown corrections to gravitational waves},\ }\href@noop {} {\bibfield  {journal} {\bibinfo  {journal} {Physical Review D}\ }\textbf {\bibinfo {volume} {101}},\ \bibinfo {pages} {084050} (\bibinfo {year} {2020})}\BibitemShut {NoStop}%
\bibitem [{\citenamefont {Isi}\ \emph {et~al.}(2017)\citenamefont {Isi}, \citenamefont {Pitkin},\ and\ \citenamefont {Weinstein}}]{lvk7isi2017probing}%
  \BibitemOpen
  \bibfield  {author} {\bibinfo {author} {\bibfnamefont {M.}~\bibnamefont {Isi}}, \bibinfo {author} {\bibfnamefont {M.}~\bibnamefont {Pitkin}},\ and\ \bibinfo {author} {\bibfnamefont {A.~J.}\ \bibnamefont {Weinstein}},\ }\bibfield  {title} {\bibinfo {title} {Probing dynamical gravity with the polarization of continuous gravitational waves},\ }\href@noop {} {\bibfield  {journal} {\bibinfo  {journal} {Physical Review D}\ }\textbf {\bibinfo {volume} {96}},\ \bibinfo {pages} {042001} (\bibinfo {year} {2017})}\BibitemShut {NoStop}%
\bibitem [{\citenamefont {Wang}\ \emph {et~al.}(2024)\citenamefont {Wang}, \citenamefont {Yang},\ and\ \citenamefont {Han}}]{lvk8wang2024tests}%
  \BibitemOpen
  \bibfield  {author} {\bibinfo {author} {\bibfnamefont {X.-L.}\ \bibnamefont {Wang}}, \bibinfo {author} {\bibfnamefont {S.-C.}\ \bibnamefont {Yang}},\ and\ \bibinfo {author} {\bibfnamefont {W.-B.}\ \bibnamefont {Han}},\ }\bibfield  {title} {\bibinfo {title} {Tests of gravitational wave propagation with ligo-virgo catalog},\ }\href@noop {} {\bibfield  {journal} {\bibinfo  {journal} {arXiv preprint arXiv:2404.14684}\ } (\bibinfo {year} {2024})}\BibitemShut {NoStop}%
\bibitem [{\citenamefont {Yunes}\ and\ \citenamefont {Pretorius}(2009)}]{lvk9yunes2009fundamental}%
  \BibitemOpen
  \bibfield  {author} {\bibinfo {author} {\bibfnamefont {N.}~\bibnamefont {Yunes}}\ and\ \bibinfo {author} {\bibfnamefont {F.}~\bibnamefont {Pretorius}},\ }\bibfield  {title} {\bibinfo {title} {Fundamental theoretical bias in gravitational wave astrophysics<? format?> and the parametrized post-einsteinian framework},\ }\href@noop {} {\bibfield  {journal} {\bibinfo  {journal} {Physical Review D—Particles, Fields, Gravitation, and Cosmology}\ }\textbf {\bibinfo {volume} {80}},\ \bibinfo {pages} {122003} (\bibinfo {year} {2009})}\BibitemShut {NoStop}%
\bibitem [{\citenamefont {Mishra}\ \emph {et~al.}(2010)\citenamefont {Mishra}, \citenamefont {Arun}, \citenamefont {Iyer},\ and\ \citenamefont {Sathyaprakash}}]{lvk10mishra2010parametrized}%
  \BibitemOpen
  \bibfield  {author} {\bibinfo {author} {\bibfnamefont {C.~K.}\ \bibnamefont {Mishra}}, \bibinfo {author} {\bibfnamefont {K.}~\bibnamefont {Arun}}, \bibinfo {author} {\bibfnamefont {B.~R.}\ \bibnamefont {Iyer}},\ and\ \bibinfo {author} {\bibfnamefont {B.~S.}\ \bibnamefont {Sathyaprakash}},\ }\bibfield  {title} {\bibinfo {title} {Parametrized tests of post-newtonian theory using advanced ligo and einstein telescope},\ }\href@noop {} {\bibfield  {journal} {\bibinfo  {journal} {Physical Review D—Particles, Fields, Gravitation, and Cosmology}\ }\textbf {\bibinfo {volume} {82}},\ \bibinfo {pages} {064010} (\bibinfo {year} {2010})}\BibitemShut {NoStop}%
\end{thebibliography}%

\end{document}